# A Knowledge base model for complex forging die machining

Bernardin Mawussi [a,b], Laurent Tapie [a,b]

[a] *Laboratoire Universitaire de Recherche en Production Automatisée ENS Cachan ; 61, avenue du Président Wilson ; 94235 Cachan cedex, France*
[b] *IUT de Saint Denis, Université Paris 13; Place du 8 mai 1945 ; 93206 Saint Denis Cedex, France*

**Abstract:**

*Recent evolutions on forging process induce more complex shape on forging die. These evolutions, combined with High Speed Machining (HSM) process of forging die lead to important increase in time for machining preparation. In this context, an original approach for generating machining process based on machining knowledge is proposed in this paper. The core of this approach is to decompose a CAD model of complex forging die in geometric features. Technological data and topological relations are aggregated to a geometric feature in order to create machining features. Technological data, such as material, surface roughness and form tolerance are defined during forging process and dies design. These data are used to choose cutting tools and machining strategies. Topological relations define relative positions between the surfaces of the die CAD model. After machining features identification cutting tools and machining strategies currently used in HSM of forging die, are associated to them in order to generate machining sequences. A machining process model is proposed to formalize the links between information imbedded in the machining features and the parameters of cutting tools and machining strategies. At last machining sequences are grouped and ordered to generate the complete die machining process. In this paper the identification of geometrical features is detailed. Geometrical features identification is based on machining knowledge formalization which is translated in the generation of maps from STL models. A map based on the contact area between cutting tools and die shape gives basic geometrical features which are connected or not according to the continuity maps. The proposed approach is illustrated by an application on an industrial study case which was accomplished as part of collaboration.*

## 1. Introduction

Design of forged part and forging die is obtained by the study of metal flow from the die cavities to the gutter and flash through the parting surface. This type of study was performed in different sections of the functional part using a wire-frame model (Bagard, 1997; Mawussi, 1995; Sun, Sequin, 2001). Then models of forged part and forging die are designed by machining assistant who is generally in charge of the definition of the final shapes and the generation of the machining process. In this approach, the final shapes of forged parts and forging dies was obtained according to the know-how of the machining assistant and the available machining resources.

Today, die designers define forged part according to the whole shape of the functional part. Therefore design of the final shapes of dies and forged parts is transferred to engineering and design department. This evolution is closely linked with the development of powerful software tools which allow simulating metal flow in the 3D model of die cavities (TRANSVALOR, Material forming simulation, http://www.transvalor.com; Kulon, Mynors, Broomhead, 2006). The die model obtained in the engineering and design department is processed by the machining assistant.

Today, die models have more complex shape. Indeed, the introduction of close die forging process (Fig. 1) increases the complexity and the finishing requirements of die shapes in order to limit machining operation (Doege, Bohnsack, 2000).

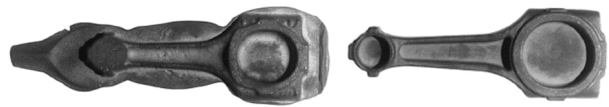

Fig. 1. Conventional versus close die forging process

Because of the evolution of die shapes, the machining assistant must now elaborate the machining process starting from die model given by the engineering and design department. This model is considered in most of the cases as a void geometry, since it involves poor technological and topological information required by machining assistant. So, machining process preparation involves the analysis of geometrical and topological information of the die shape in order to associate machining information. At this step of the elaboration of the machining process, the machining assistant chooses machining strategy and resources according to die shape with respect to the minimal machining time, the part quality and surface roughness. Integration of High Speed Machining (HSM) technology is also crucial for the processing of technical and economical constrains.

In the die machining context, HSM does not only mean high cutting speeds and/or high feed rates. It should be considered as a machining process where operations are realised with specific methods and equipment (machining strategies, machine tool, cutting tools, spindle…) and involves machining process modifications (SANDVIK Coromant, Die & Mould Making, Application Guide,

http://www.coromant.sandvik.com, 2010). Electro Discharged Machining (EDM) step is significantly reduced and often eliminated from die machining process. Heat treatment after die semi finishing machining can be avoided thanks to improvement in cutting tools manufacturing and preparation. As the machining of forging die is performed on high heat treated stocks (45-60 Hrc) several technical and economical reports highlight productivity and part quality increases with the introduction of HSM (Kranjnik, Kopac, 2004; Toh, 2005; Altan, Lilly, Yen, 2001). These evolutions must lead to a better part quality for HSM than conventional machining if an efficient machining planning is associated to a relevant choice of cutting conditions and tools. In this context the tasks carried out are very difficult to automate owing to the complexity of the die shape and involve high skilled worker intervention realising time expensive operations.

Current forging die machining process is decomposed in three main steps: roughing, semi-finishing and finishing. Roughing operation objective is to reach as close as possible the desired rough shape with high level of cutting material flow and tool path series as smooth as possible. Associated machining features for this operation are pockets or cavities. Semi-finishing operation objective is to warrant a constant depth of cut or constant tool engagement during finishing operation. Finishing operation objective is to obtain the final die shape up to the geometrical quality specified. The die geometry and topology must be taken into account during this final operation with a high levelled refinement to reach this objective.

The work presented in this paper is devoted to finishing operation of die cavity. According to the industrial context, the finishing process is based on standard machining resources (3-axes high speed machine tools, end mill, ball nose and corner rounded end mill cutting tools) and common Computer Aided Machining (CAM) strategies (parallel planes, Z level, parallel curve guidance). These common CAM strategies presented in several works (Toh, 2005; Dragomatz, Mann, 1997) are integrated in a very wide CAM and Computer Aided Process Planning (CAPP) software.

Technical and economical studies highlighted the importance of CAM preparation in the die machining process. During this preparation step, the experience and know how of the machining assistant has a great influence on the part quality. Thus, the tasks carried out must be adapted to this technical and economical context which involves evolutions on CAD models, CAM strategies and machining strategies (HSM). This adaptation depends mainly on the good suitability between die shape and HSM process.

CAM preparation is based on machining feature recognition from the CAD model of the part. Several approaches for machining features recognition are proposed in different works: syntactic pattern recognition (Abouel Nasr, Kamrani, 2006; Sundararajan, Wright, 2004), graph-based recognition (Lee, Jhee, Park, 2007), rule-based recognition (Choi, Ko, 2003; Zhang et al. 2004; Sridharan, Shah, 2005). These approaches are based on a well known machining feature data base and a topological analysis of the CAD model geometry. Yet, the topological analysis of the CAD model must be closely link with the machining process specificities. (Sunil, Pande, 2008) highlight that efficient machining feature recognition is always carried out by the formalisation of knowledge used by machining process expert. (Sun, Sequin, Wright, 2001) propose 3 axis machining feature recognition based on cutting tool tip geometry and topological analysis of the slope angle of the part. Some approaches are developed for 3-5 axes HSM of aeronautical structural parts illustrated in Fig. 2.a (Derigent et al., 2007, 2010; Yu et al., 2008). Topological analysis of shapes denotes that forging dies are composed of several cavities having very different shapes and depths Fig. 2.b compared to pockets of aeronautical parts. Beyond this difference, no machining features data base is formalised in industrial practices for forging die machining even though for aeronautical parts machining features recognition is based on a well-known machining features data base.

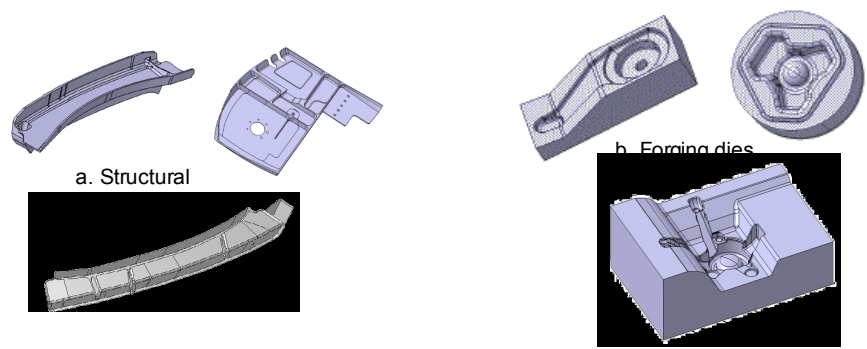

Fig. 2. Complex shaped parts with cavities

The work presented in this paper is aimed to transform a complex shaped CAD model to a CAM model towards an efficient identification of machining features. The main objective of the approach presented is to define machining features suitable for HSM process from the die model (3D CAD model). First, a machining process model integrating the machining feature concept is presented. Then, based on this model the knowledge useful for the machining feature identification on forging die CAD model are detailed.

## 2. Machining process

Traditional machining process generation is conducted towards four main steps Fig. 3. In the first step, the CAD model is analyzed in order to collect geometrical, topological and technological information (Fig. 3 step 1). The results of this analysis are associated and then used to create geometrical features. This step states the decomposition of the part into several geometrical features. In the second step machining strategies and cutting tools are selected for each geometrical feature (Fig. 3 step 2). The parameters of these resources are defined based on the analysis carried out in the first step. Several CAM software allow computing tool paths. According to the machining strategies and cutting tools parameters defined in the second step, the appropriate module is selected from the CAM software in order to generate machining sequences (Fig. 3 step 3). At last, machining sequences are grouped and planed in order to obtain the machining process (Fig. 3 step 4). This planning process is also based on the results of the topology analysis.

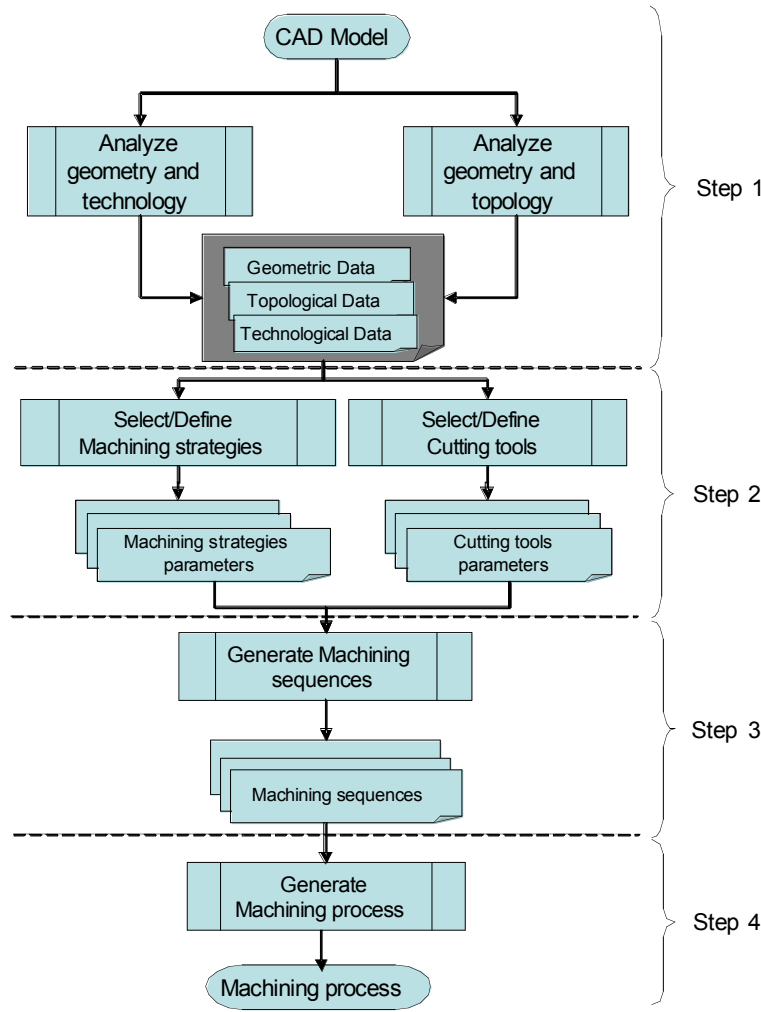

Fig. 3. Machining process generation

This paper is focused on the geometrical and technological analysis of the CAD model. To ensure the completeness of the geometrical and technological analysis, a machining process model which is structured towards the four main steps presented in Fig. 3 is proposed. Knowledge carried out in step 2 are formalized in the following in order to link the different components of the machining process model given at the last part of this section. The new approach developed on the basis of the knowledge formalized in this paper allows extracting machining features for step 3.

### 2.1 Geometrical and technological analyses

To generate a machining process by using any CAM software, it is necessary to identify the surface or the portion of surface to be machined and to specify the machining limits on the CAD model of the part. This task is performed starting from a geometrical data analysis with respect to the technological data defined by the designer. Typically, information on material, form tolerance and roughness defined as functional requirements by the engineering department are considered as a part of technological data. Currently, surface to be machined does not correspond to all the surface of the part. The machining process must be adapted in order to satisfy the cutting conditions which depend on the technological data.

The machining data such as the scallop height and the machining tolerance defined in this level starting from the technological data are associated with surface to be machined in order to create a machining feature (Fig. 4). A machining feature is defined as the set made up of a geometrical feature and its technological data to which it is possible to associate a well-known machining process. A similar definition was given in (Bourdet et al., 1990) for prismatic parts.

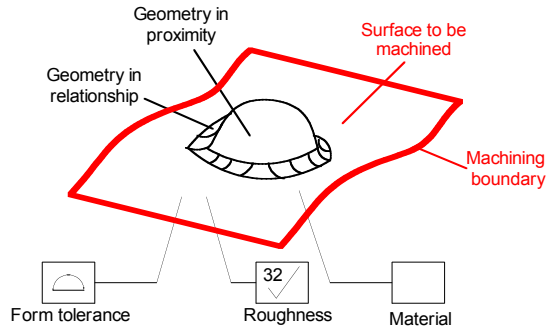

Fig. 4. Machining feature

When the machining feature is created, the tool paths are computed by considering the other geometrical shapes located in its environment. Analysis of the geometry and the topology of the part thus make it possible to avoid interferences with the part in the zones where the material should not be removed.

## 2.2    Geometrical and topological analyses

During machining process generation, it is necessary to analyse the geometry of the part according to several topological data. The different types of topological and geometrical data considered allow distinguishing the geometrical shapes in contact and the proximity relation of the surface to be machined (Fig. 4).

All the topological relationships are not detailed in this paper. The examples presented on figure 5 show the three main situations in which the computation of the tools paths takes account of the topological relations and the type of shapes with which they are associated. In Fig. 5.a, the surface to be machined has topological relations with concave surfaces. The cutting tool can move along the initial surface without any deviation of the tool paths. In Fig. 5.b the surface to be machined has topological relations with convex surfaces. The cutting tool cannot move entirely along the initial surface. The tool path must be modified to avoid interference with the convex surfaces. In Fig. 5.c the surface to be machined has a contact topological relation with a concave surface and a proximity topological relation with a convex surface. The cutting tool can move along the initial surface on the top of the concave surface but the tool paths must be modified to avoid interference with the convex surface.

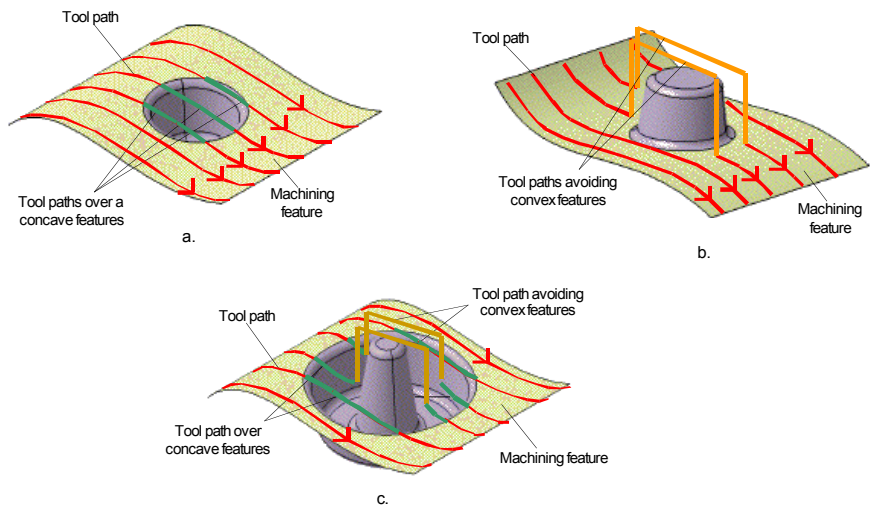

Fig. 5. Influence of topology on tool paths

To integrate all the topological situations in the machining process generation it is necessary to build the complete model of the topological relations. This is done in our work and the results will be presented further. When the two phases of the geometrical analysis are completed the tool paths can be computed thanks to the selection of cutting tools and machining strategies.

### 2.3 Machining strategy

To define a machining strategy, four parameters must be set. The first parameter is the feed direction which corresponds to the sweeping direction of the machining feature by the cutting tool. Three types of feed direction can be defined. In Fig. 6.a the direction is given by a curve to which the tool paths are parallel. When this type of strategy is selected, the difficulty is to determine the curve which makes it possible to obtain a most constant feed rate as possible. The solution consists in extracting the curve starting from the geometry of the machining feature. For the two other types of direction, the tools paths are located in plans parallel with that selected for the strategy. This plan can contain the tool axis or be perpendicular to it Fig. 6.b. The sweeping mode gives an additional detail on how the transition is made between the outward paths and the return paths Fig. 6.c.

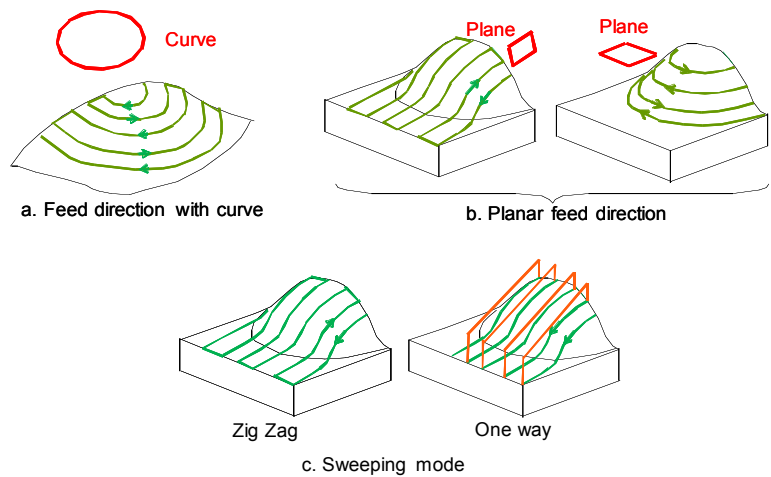

Fig. 6. Machining types and parameters

The second parameter of the machining strategy is the cutting mode which defines the correspondence between the direction of rotation of the cutting tool and that of its feed rate Fig. 7.a. Two modes can be defined: downward and upward. When the upward mode is chosen, the teeth of the cutting tool in contact with the material advance in opposite direction of the machining direction. So they start with the maximal chip thickness and finish with a null chip thickness. The downward mode is the reverse of the upward mode. For the machining of the complex shapes and thus of our machining features, the parameter is set to the upward mode because it facilitates the cutting conditions.

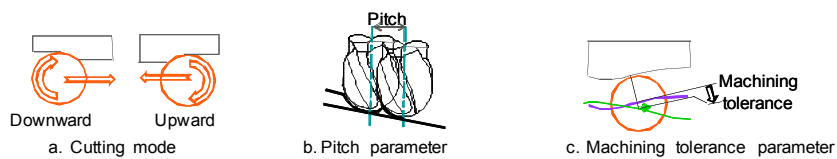

Fig. 7. Machining quality requirements

The two last parameters are linked to the dimensional quality of the surface machined. The pitch parameter which represents the radial depth of cut defines the scallop height (Fig. 7.b). The scallop height defined in CAM tools used in our work is directly linked to the surface roughness (Rt criteria). If the criteria defined in the technological data are different from the Rt, tables available in the literature to establish conversion are used.

The machining tolerance parameter defines the maximal deviation between the tool path and the machined surface (Fig. 7.c.). It is directly declared by transposing the form tolerance defined in the technological data.

Once the parameters of the machining strategy are defined, it is necessary to associate them with the different elements of the machining feature in order to integrate the results of the analyses into the machining preparation tasks. The structure of the associations is presented in Fig. 8. Only the associations of the machined surface with the sweeping mode and the feed direction are detailed in section 3. Associations relating to the topological relations and the machining boundary which are also important will be developed in another communication.

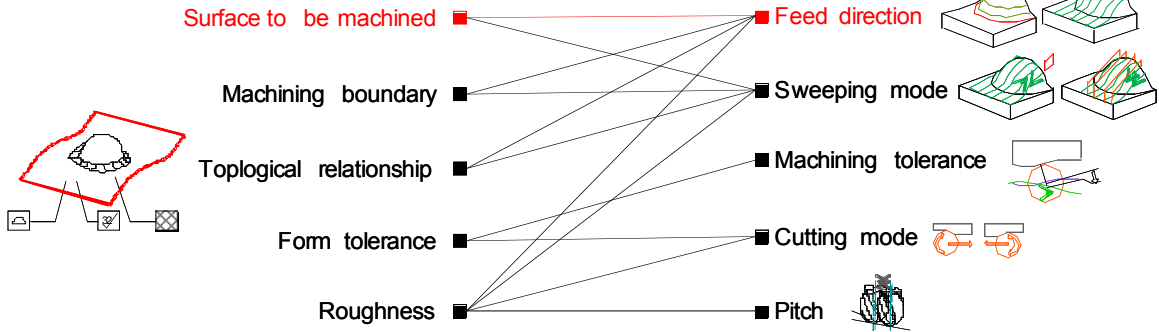

Fig. 8. Machining features and strategies links

## 2.4    Cutting tools

Selection of cutting tools follows the same steps of the machining strategies. It is based on the parameters which define the tool tip and the geometry of its body and attachment which is necessary to check the accessibility of the machined surfaces. From the APT model considered for the definition of the tool tip (Fig. 9a) three types of tools used generally in the machining processes of complex shapes parts are identified (Fig. 9b). Only two types "ball nose" and "corner end mill" are used in finishing operations. The flat end mill is generally used in roughing operations but it can be appropriate for the finishing of some convex shapes with plane surfaces. To avoid the interference between the tool and the part, the dimensions OL and TL (Fig. 9c) are checked for their compatibility with the depth of the cavities. The material of the cutting tool is also an important parameter. It is directly determined from the material of the part.

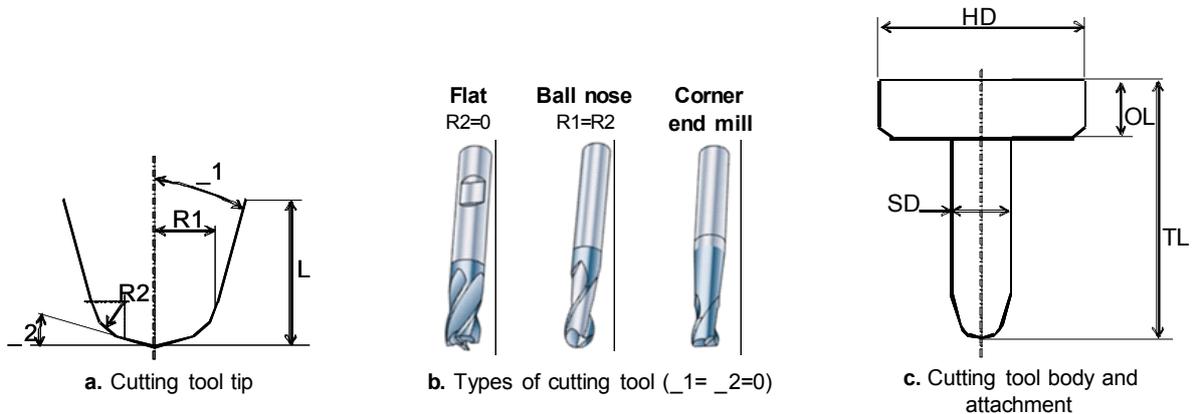

Fig. 9. Cutting tool parameters

Associations of the cutting tool parameters with the different elements of the machining feature are presented in Fig. 10. The relation between the roughness and the tool tip corresponds directly to that which was established with the pitch parameter of the machining strategy (Fig. 8). Indeed, the geometry of the tool tip determines the scallop height according to the pitch. The tool tip can be more finely defined according to the machining boundary. As specified here before, this association will not be detailed in this paper. The determination of the tool will thus be done through its type.

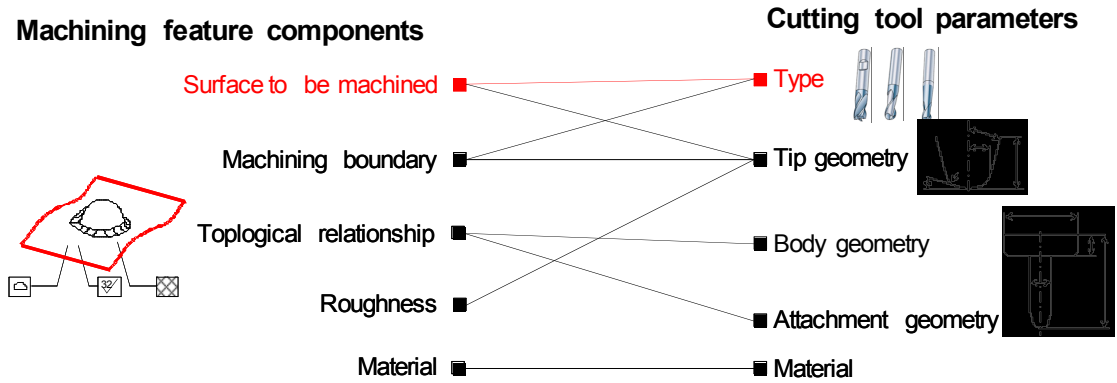

Fig. 10. Machining features and cutting tool links

## 2.5 *Machining sequence*

Tool paths can be generated when the machining strategies are defined and the types of cutting tools are selected. For a given machining feature, the machining sequence is aimed to gather its tool paths. It is thus defined as an uninterrupted and organised series of machining trajectories computed with a unique cutting tool, a unique approach and a unique clearance (Fig. 11). Through this organization, a given machining sequence is composed of (in this order) an initial trajectory, one or several intermediate trajectories and a final trajectory. Each trajectory is also composed of an approach path, a machining path and a clearance path. In addition, the initial and final trajectories have respectively a fast initial approach and a fast final clearance.

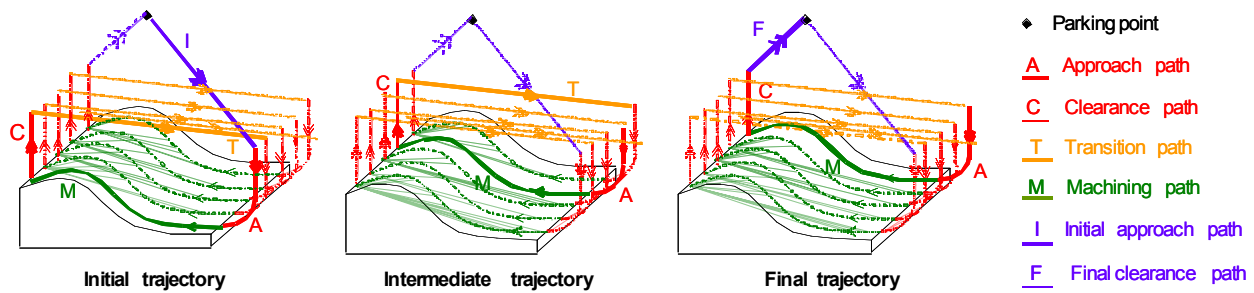

Fig. 11. Machining sequence components

The most important part of a machining sequence is the set of machining tool paths along which the cutting conditions must be respected in order to obtain the final part. Then, parameters chosen or defined for the cutting tools and the machining strategies must be compatible with the performances of the machine tool in terms of axis kinematics and Numerical Controlled Unit (NCU) capacities. The example presented in Fig. 12 shows some problems related to this compatibility. This result was obtained by using "performance viewer" which is a simulation tool developed during our research works (Tapie, Mawussi, Anselmetti, 2007a, 2007b). It highlights feed rate reductions: the set point of the feed rate is not respected or reached. The significant feed rate reduction 2 is induced by the small radius of curvature of the portion of the surface to be machined. Feed rate reductions 4 and 5 are induced by the topological relationship between the machined surface and two limit surfaces. Finally, the feed rate reduction 1 is also related to a topological relation but as the surface is convex, the tool paths can be extended in order to limit this reduction.

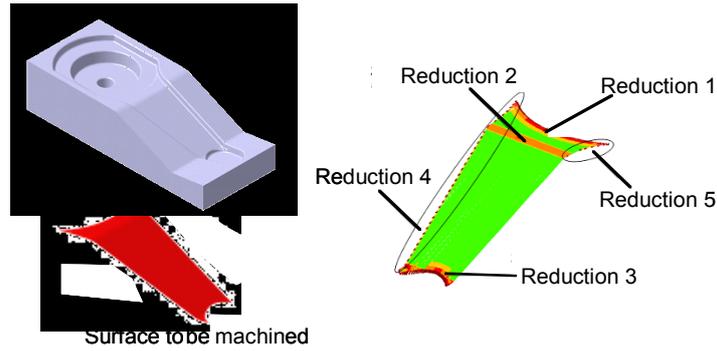

Reduction 2
Reduction 1
Reduction 4
Reduction 5
Reduction 3
Surface to be machined

Fig. 12. Feed rate reductions

The generation of the machining sequences is carried out thanks to the association of the machining strategies and the cutting tools to the machining features. The association process (Fig. 13) comprises five tasks. Approach/initial paths and clearance/final paths are respectively generated during tasks 1 and 2. Their basic shape depends in most cases on the body geometry of the cutting tool, the machining strategy (sweeping mode and feed direction) and the surface to be machined of the machining feature. This basic shape must be adapted according to the attachment geometry of the cutting tool and the topological relationships of the machining feature in order to avoid collisions. Transition paths are generated in the same way during task 3. The cutting mode and the pitch parameter determine which machining paths have to be linked up by transition paths and how. The generation of these machining paths during task 4 is mainly based on the tip geometry of the cutting tool and all elements of the machining strategy. As the parking point must allow avoiding collisions, it is defined during task 5 from the body and attachment geometries of the cutting tool and the topological relationships of the machining feature.

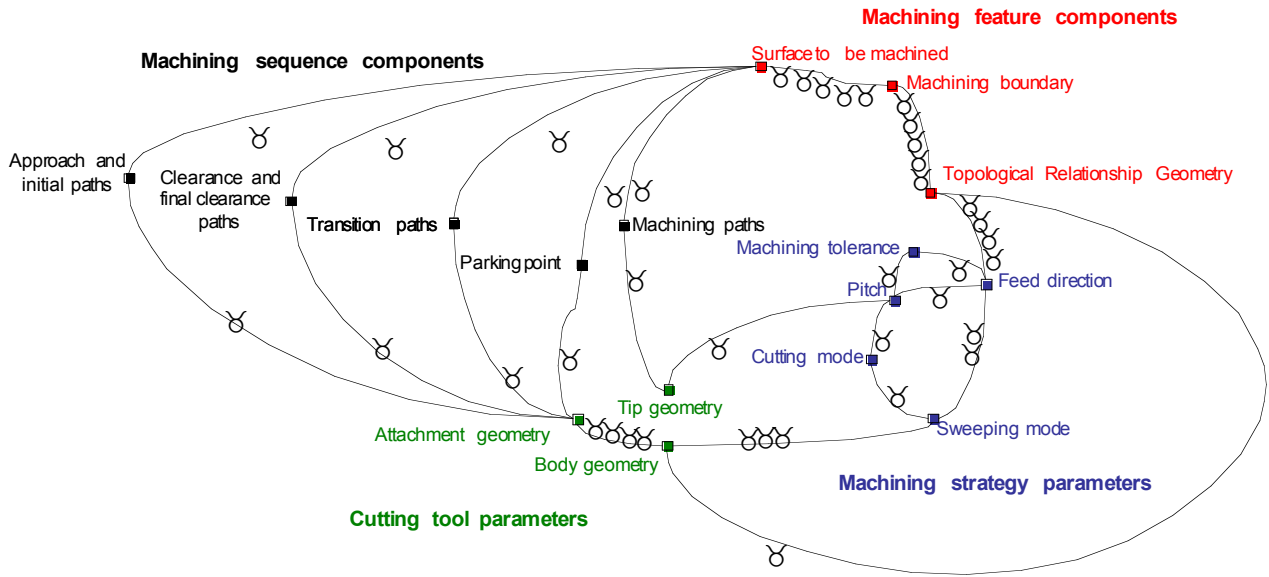

Fig. 13. Links between machining process components

### 2.6    *Machining process model*

The generation of the machining process is made from the ordered gathering of the machining sequences. Data used during this generation are linked ones to the others. These links have been modelled using UML language (Fig. 14). Clustering of topological and technological data to the basic geometric shape in order to create the machining feature is presented in the left of the data structure. Then the association of the cutting tools parameters and the machining strategies elements in parallel to each machining feature created previously make it possible to generate the machining sequence. Finally all machining sequences are clustered in the machining process of the part towards a planning task which takes into account topological relationships between the machining features and of which objective is to reduce the machining time. In the following part only the knowledge formalization during the generation of the machining sequences is presented.

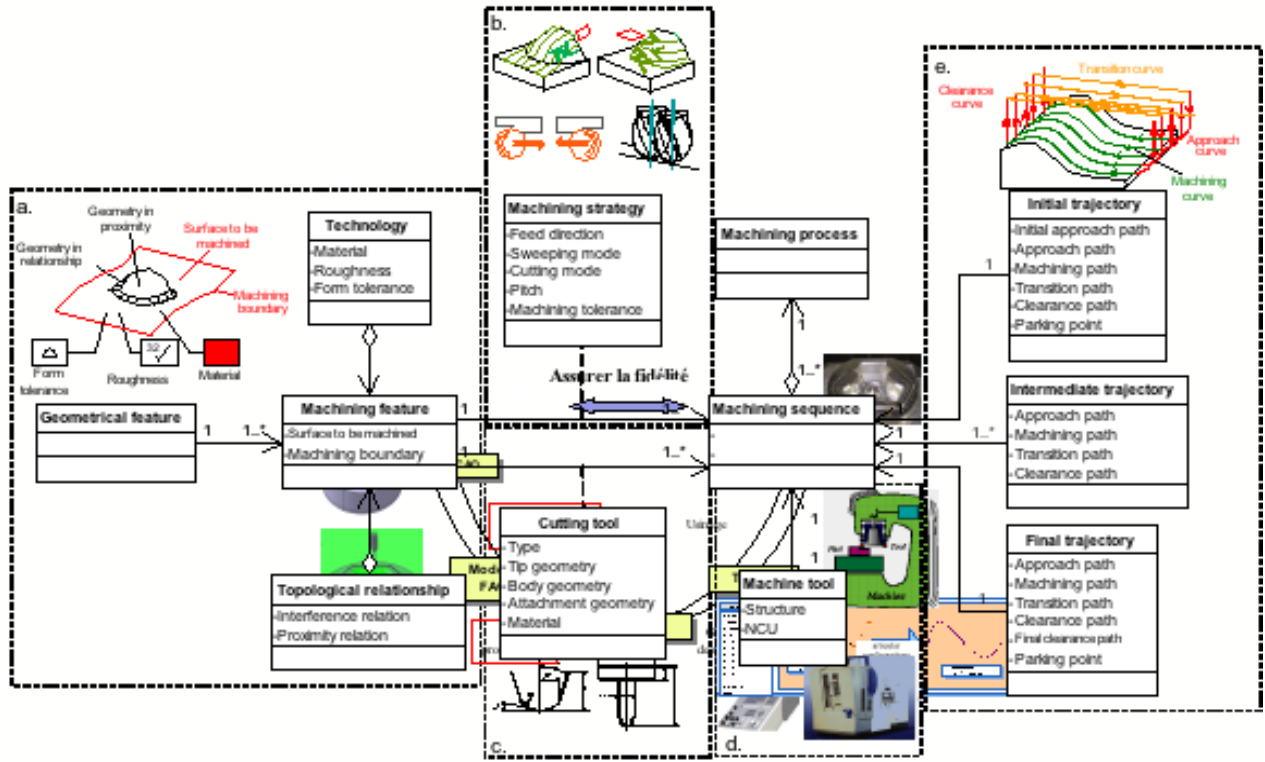

Fig. 14. Machining process model

## 3. Knowledge integration

Extraction of the different geometrical features starting from CAD model of the part is an important task insofar as it must make it possible to isolate areas of the part to which are associated the cutting tools and the machining strategies. Therefore the approach developed in this paper is aimed to integrate the know-how of the machining assistant in the decomposition of the part shape into geometrical features.

In industrial practices of forging die machining, no machining features data base is formalised. Indeed, the CAD model is not sufficient to associate a well known machining process to a part area. The difficulty is to identify a machining feature (in particular the surface to be machined) suitable with a machining process.

The proposed approach is to associate machining data to the part surfaces in order to create machining features. The part geometry is analyzed according to machining criteria in order to define surfaces to be machined (Fig. 15). Two machining criteria are mainly developed in this section: cutting tool type and machining strategy.

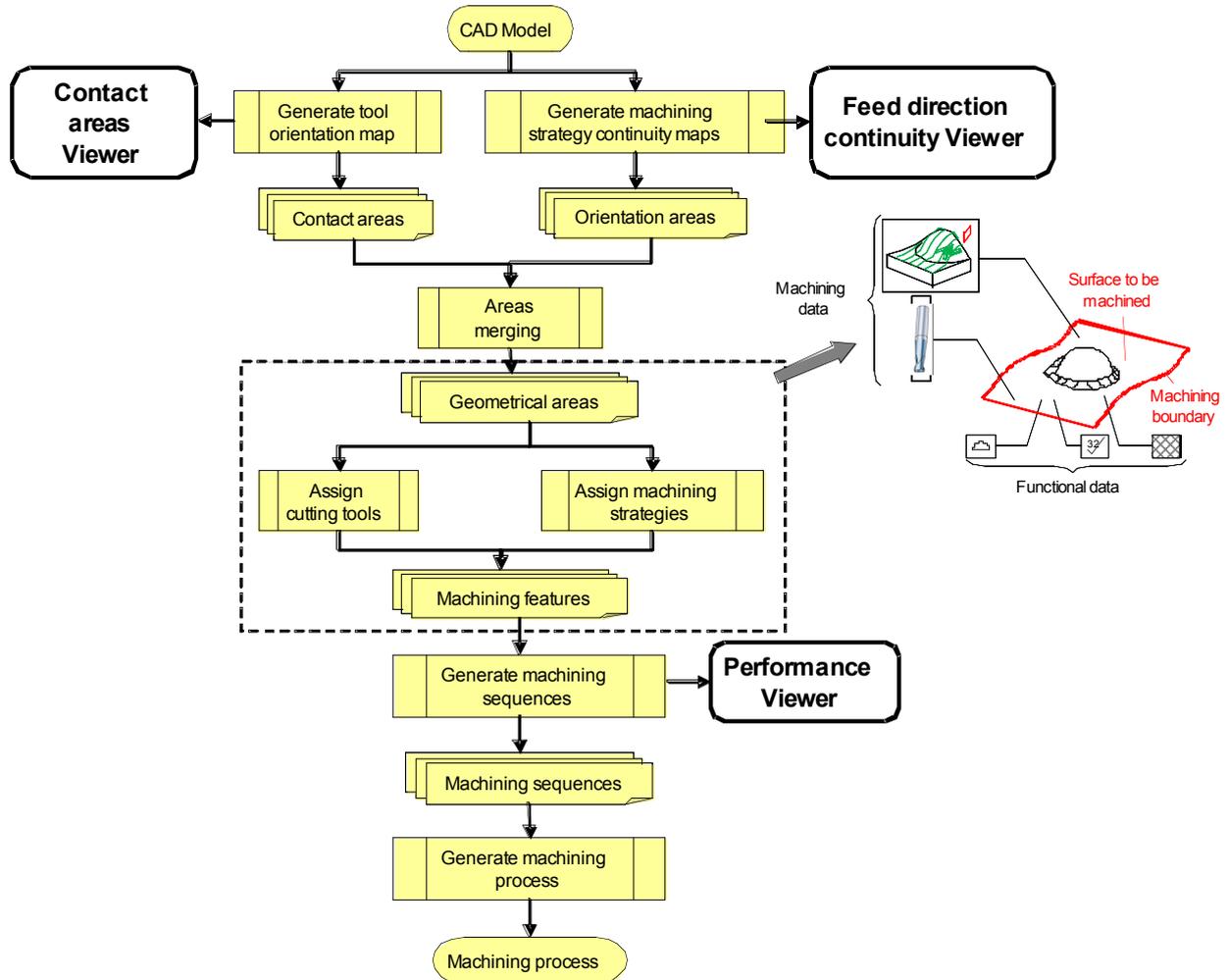

Fig. 15. Machining process generation with knowledge integration

## 3.1 Contact area on cutting tools

One of the main difficulties during high speed machining of complex shape parts is to maintain the cutting speed set point because the contact area between cutting tool and the shape of the part changes every time (Fig. 16). Maintaining the cutting speed is a key factor in the cutting mechanism (Tapie, Mawussi, 2008). This machining difficulty is closely linked with the tool type and the geometry of the part. Indeed, to machine a horizontal plane and to maintain the cutting speed, a flat or corner end mill is more appropriate than a ball nose end mill.

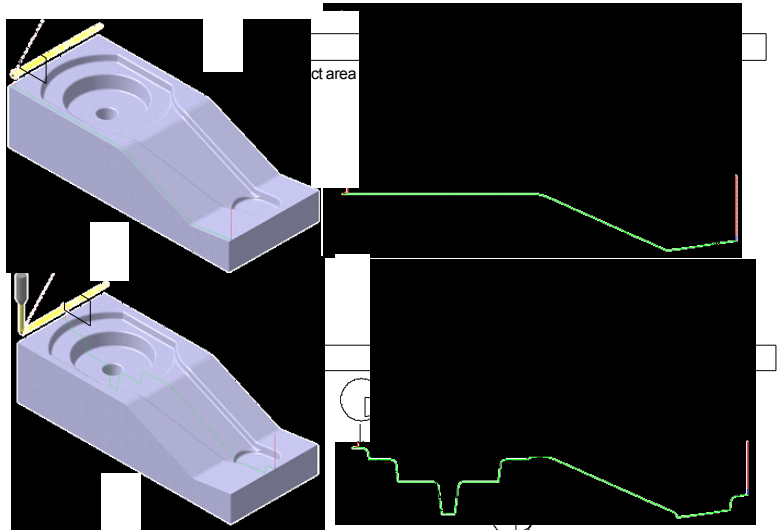

Fig. 16. Examples of contact areas

The orientation of a given facet of the STL model of the die compared to the tool axis (or machining direction) makes it possible to locate the contact area between tool tip and the machined surface and to estimate its size. Two types of significant information can be extracted from the analysis of this orientation: indication about the range of the contact area (closely linked with the area size) and evaluation of the effective radius of the tool tip which determines the effective cutting speed. The principle of building the contact map is based on the evaluation of the tool orientation obtained by the projection of the unite vector representing the STL facet normal on the tool axis direction. The orientation indicator resulting from projection (**Erreur ! Source du renvoi introuvable..**a) defines the location of the contact area between tool and part. Indeed, when the contact area moves from the center of a ball en mill tool to its external diameter, the orientation indicator takes a value which varies from 0 to 1. Inversely to the orientation indicator, the projection of the unite vector representing the STL facet normal on a plane H perpendicular to the tool axis direction defines a contact area indicator (**Erreur ! Source du renvoi introuvable..**a) which represents the size of the contact area. Orientation or contact area map is generated according to the two indicators highlighted previously. The map presented in **Erreur ! Source du renvoi introuvable..**b is obtained by assigning a color to the contact area indicator of each STL facet. The color scale is the one provided by MATLAB[®]. Analyzing the results obtained from maps generated for several forging dies, three types of contact areas can be identified (**Erreur ! Source du renvoi introuvable..**c). Flat contact area, which is the first type, characterizes planar or quasi planar surface portions. Usually this type of area defines cavity bottoms and die parting planes. Draft area, which is the second type, characterizes drafted surfaces (vertical or quasi vertical) representing cavity flanks. Between the colors of the two first types of contact area, several surface portions are associated to graduated colors. The transition contact areas associated to these colors characterize transition or blend surfaces located between flat and drafted surfaces.

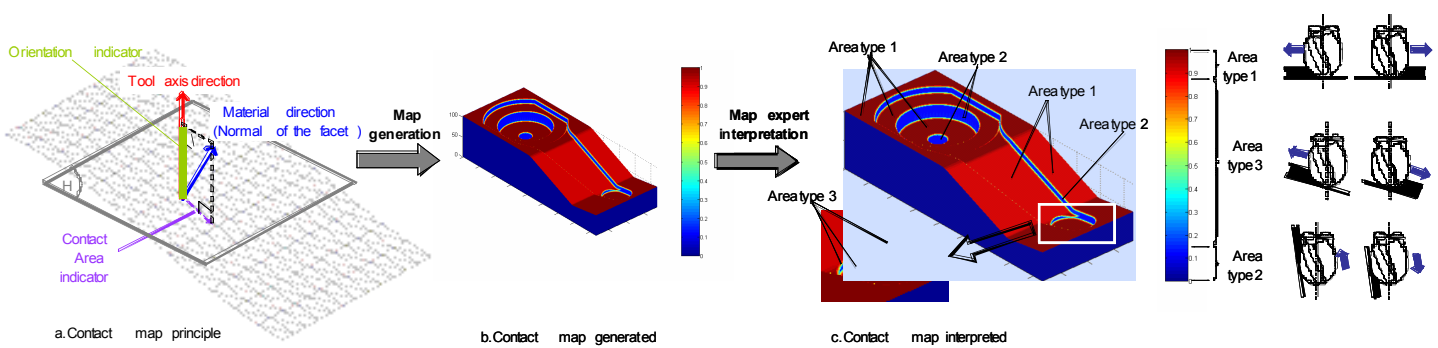

Fig. 17. Elementary geometrical feature identification

Thresholds of the color limits between the different types of area are fixed by the machining assistant according to technological data. In the map shown in **Erreur ! Source du renvoi introuvable..**b they are fixed to 0.15 for the draft contact area and 0.8 for the flat contact area. These values can be adjusted by the machining assistant according to the machining strategies used to generate tool paths. Contact areas extracted at this level of the decomposition process give information useful for associating cutting tools to elementary geometrical features and to predict the effective cutting speed. Yet, this information is not sufficient to choose the machining strategy. A second map based on the analysis of the machining continuity is developed for the final choice of machining strategies.

### 3.2 Continuity in machining feed direction

As it is depicted previously, the choice of a machining strategy for machining tool path generation has two main objectives: avoid significant federate loss and reducing as possible machining time. To achieve these goals, the machining assistant often associates to a standard 3 axes HSM two types of machining strategies widely used in CAM software for the machining of forging dies:
- parallel planes strategy : the machining feed direction is defined by parallel planes containing the tool axis;
- Z-level strategy: the machining feed direction is defined in parallel planes perpendicular to the tool axis.

Several machining experimental works on inclined planes (Kang et al., 2001; Kecelj et al., 2004; Toh, 2004a, 2004b) or complex shape parts (Baptista, Antune Simões, 2000; Vivancos, 2004) have been performed. Analysis of the results of these experimentations underlines the fact that geometrical features must be grouped according to the same direction sequences - equivalent to the feed direction - in order to limit cutting mechanical deteriorations and thermal problems. Parallel planes strategy allows reducing machining time by covering large areas of machined parts. Priority is given to the application of this strategy towards the second type of map developed. The principle of this second map belongs to the ability of the normal vector of different adjacent facets to hold on the same plane along a machining tool path. This ability can be evaluated from the projection of the contact area indicator in the feed direction (**Erreur ! Source du renvoi introuvable.**a.) which characterizes the continuity indicator of parallel planes strategy. Indeed, when the normal vectors of several adjacent facets hold on the same plane containing the tool axis direction and the feed direction, their projections on the plane H must be equal to the contact area indicator defined previously (see **Erreur ! Source du renvoi introuvable.**a.). If the continuity for parallel planes is not obtained for several adjacent facets but the values of the contact area indicators are the same, the resulting machining condition corresponds to continuity in the plane H which becomes the Z-level plane. For this reason the contact area indicator becomes the Z-level indicator in this stage (**Erreur ! Source du renvoi introuvable.**a.). This Z-level indicator characterizes the ability of the normal vectors of different adjacent facets to turn around the feed direction with the same contact area. Then, the contact map generated previously also allows determining the Z-level continuity for the final geometric features extracted.

Analyzing the results obtained from maps generated for several forging dies, 3 machining continuity cases can be distinguished:
- Indifferent feed direction (**Erreur ! Source du renvoi introuvable.**b.): it corresponds to flat areas for which the normal vectors of several adjacent facets are hold on a same plane whatever is the selected feed direction.
- Oriented feed direction (**Erreur ! Source du renvoi introuvable.**c.): it corresponds to the continuity of planar or quasi-planar areas for which only one feed direction is found.
- Undefined feed direction (**Erreur ! Source du renvoi introuvable.**d.):  it corresponds to transition areas or blend surfaces for which there is no privileged feed direction.

A fourth case can be highlighted. It corresponds to draft surfaces for which the Z-level strategy continuity indicator is constant in several adjacent facets. This case is equivalent to the draft area identification from the contact area map.

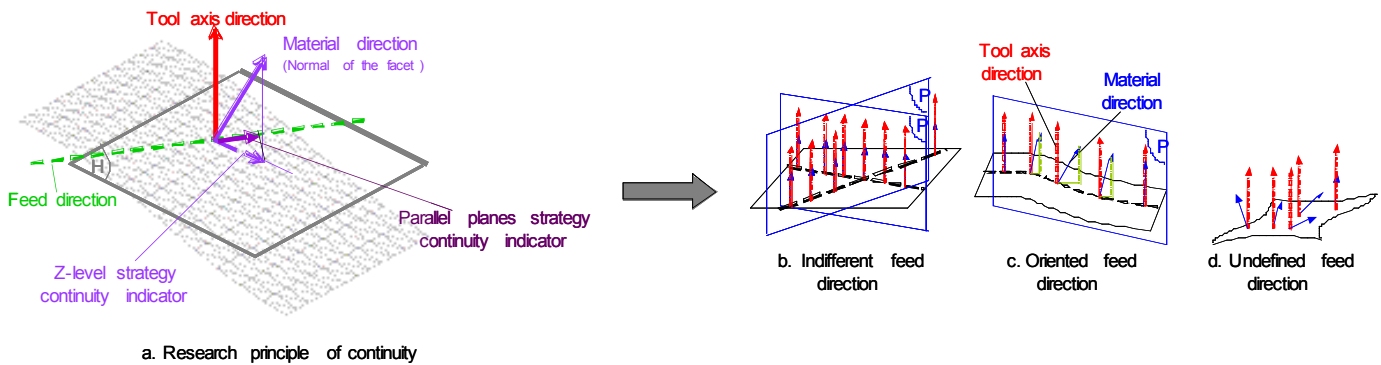

Fig. 18. Continuity in machining feed direction

In practice continuities in parallel planes are searched by testing several feed directions. Ten directions are systematically tested between 0 ° and 90 ° with a step of 10 °. In the example of **Erreur ! Source du renvoi introuvable.**, only one oriented feed direction is identified on the first map. To be efficient, tested directions must be determined from the characteristics of the contact areas and the main directions of the rough shape of the part. Variations of the feed direction around these directions can be small.

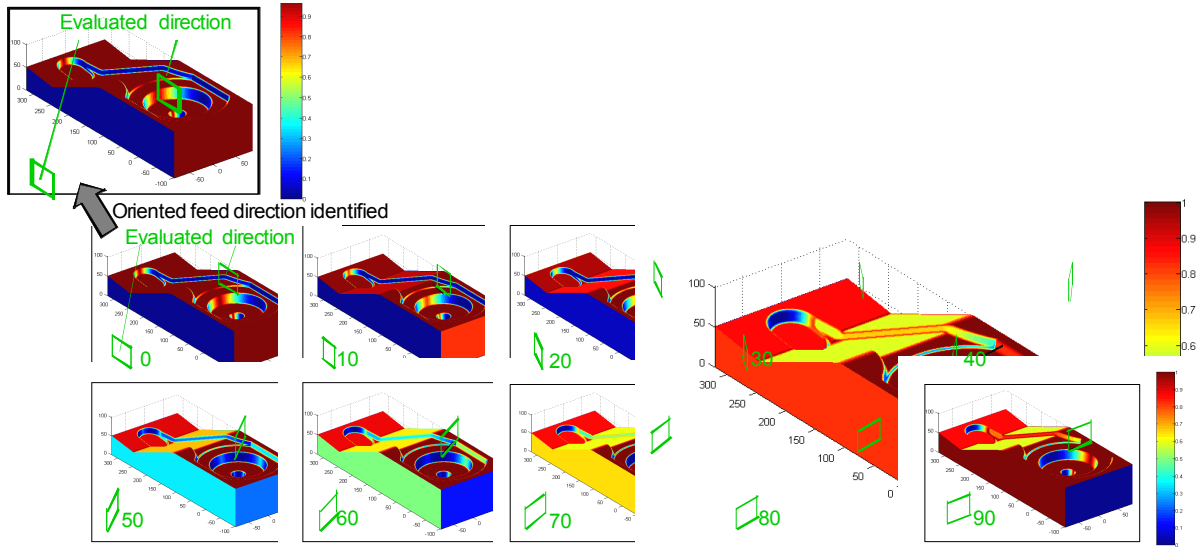

Fig. 19. Machining continuity maps

Geometrical features are finally identified from areas obtained by bringing closer maps corresponding to tool orientation and machining feed direction continuity. The maps combination is carried out toward their complementarities. Indeed, the tool map orientation, realised first, integrates knowledge about cutting tool choice and cutting conditions and enables to obtain elementary geometrical features. The machining strategy continuity maps, built in a second stage, enable to group or not elementary geometric features and then integrate relevant knowledge for machining strategy choice.

Application introduced afterwards shows a complete identification of areas as well as different associations of cutting tools and machining strategies.

## 4. Application

The proposed method is applied to the generation of the machining process of the industrial die of which the CAD model is presented in Fig. 20.a. This die is composed of a deep and complex shape cavity which has a topological relationship with a central slot. The STL model (Fig. 20.b.) used to build maps is created thanks to a standard meshing tool of the CAD software CATIA®. With basic parameters (deviation of 0,2mm and maximum edge length of 1mm) the meshing process leads to 13450 facets. The accuracy of the geometric features extraction can be improved by increasing the number of facets thanks to the modification of the parameters but it will also increase the processing time. Since this improvement does not have influence on the maps, the initial number of facets is maintained.

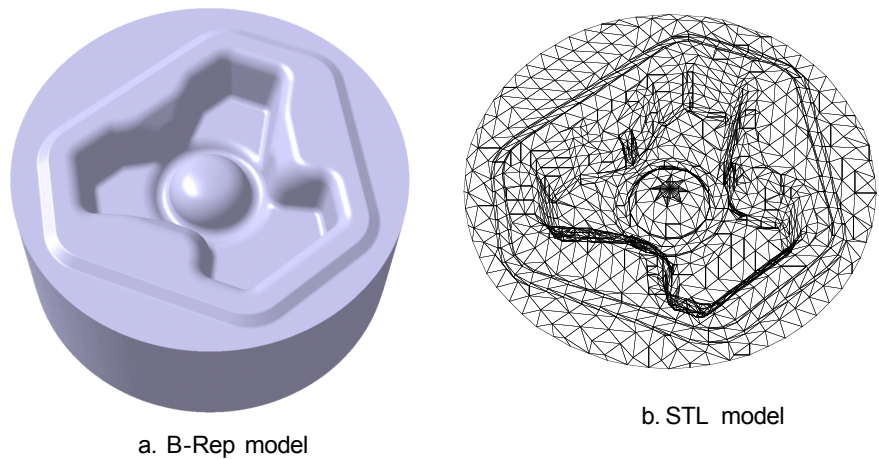

a. B-Rep model          b. STL model

Fig. 20. Industrial case study

Two maps are generated for the industrial die from the STL model (Fig. 21). The first map associated with the cutting speed shows three types of areas (Fig. 21a). Areas of types 1 which are located at the centre of the cutting tool spread over the broad part of the die. Areas of type 2 located on the flank of the cutting tool are linked up to the first by intermediate areas of type 3. The second map which allows searching for directions pointed out for a sequence of several surfaces is generated according to nine directions. For this industrial die, this map brings any other relevant information for the decomposition of geometric shape.

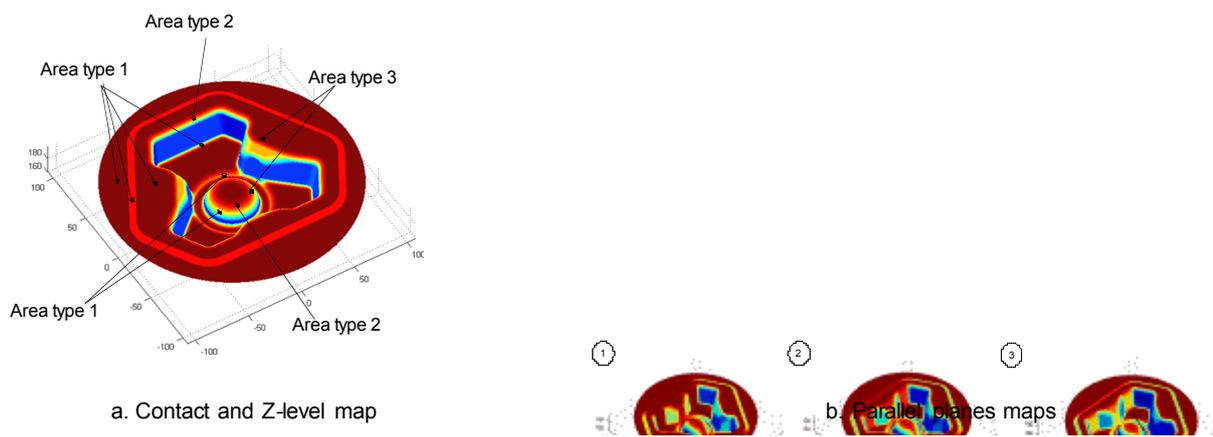

a. Contact and Z-level map

b. Parallel planes maps

Fig. 21. Maps generation

Analysis of the types of areas and their relative positions (topology) allows us to define the geometric features presented in Fig. 22. The main area of type 1 located on the top of the die represents its parting plane. The die cavity is composed of three main features: the flank surface and the bottom surface which includes a protrusion. The protrusion includes several contact areas with the cutting tool and it is not easily identifiable as much as it appears on the map through several colours. This is the main element for the recognition of revolution features composed of a rounded crest and several rings. The protrusion is linked to the bottom of the die cavity by transition features. It is also the case of the die cavity which is entirely linked to the parting plane by transition features. The height of the protrusion is such as its rounded crest does not exceed the parting plane which can therefore be machining without taking into account the protrusion. This information expresses the importance of topological relations. As a result, each machining feature is created from a form feature which embeds technological data (material, roughness, form tolerance) and topological data (example: "the protrusion does not exceed the parting plane").

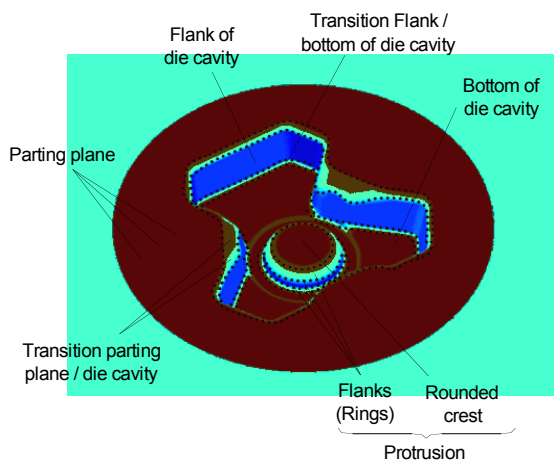

Fig. 22. Geometric features extraction

At this stage of the generation of the machining process of the die, machining features are entirely identified. Cutting tools and machining strategies associated to them according to knowledge defined in the maps (**Erreur ! Source du renvoi introuvable.** and **Erreur ! Source du renvoi introuvable.**) and methods presented in the association graphs (Fig. 8 and Fig. 10) are shown in Fig. 23. Two cutting tools and two machining strategies are necessary and sufficient for all machining features of the die. As the protrusion feature is particular, it is machined by two cutting tools because the roughness allows it (16 µm). If the junction between the rounded crest and the flank (the set of rings) of the protrusion feature owed to the change of cutting tool is not compatible with the form tolerance (0.1 mm), it is necessary to consider the use of the only corner end mill. Machining with the only corner end mill will make it possible to reach the set point of the cutting speed on the rounded crest but will increase the machining time. It will also fix only one machining strategy which will be Z-level.

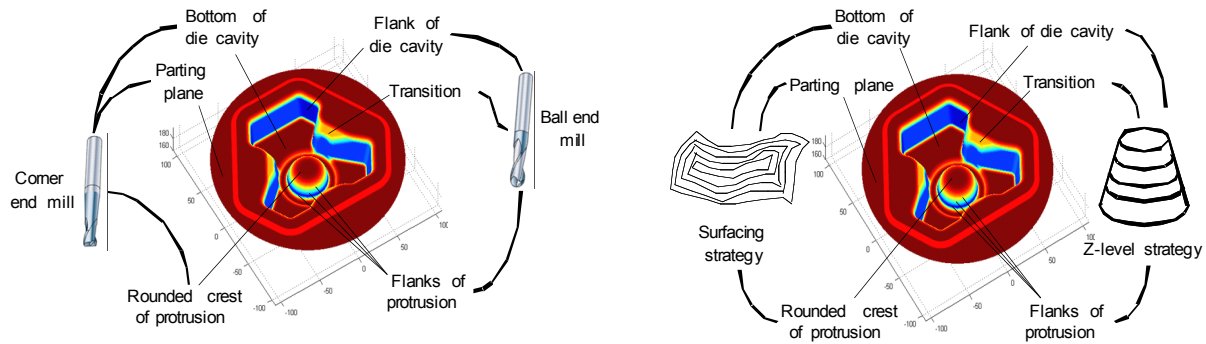

Fig. 23. Association of cutting tools and machining strategies

## 5. Conclusions

Application of a unique standard machining strategy and a choice of one cutting tool for the machining of a whole surface of a given forging die are generally unsatisfactory because of the variations of cutting speed and feed rate. Approach presented in this paper is aimed to decompose the die geometry in several machining area in order to limit the variations of cutting speed and feed rate. The decomposition process represents the CAD/CAM integration in the field of the machining of forging dies and it can be extended to complex shape parts.

The generation of the machining process of dies by the machining assistant is based on several expert knowledge which are linked to technological and topological data, machining strategies, cutting tools and machining sequence components. The detailed analysis of these expert knowledge made it possible to highlight their relations and associations to the different information being able to be processed during the decomposition of the forging die shape. The data handled at this stage as their relations were gathered in the machining process model which was used as a basis for machining knowledge formalization and integration.

Knowledge formalization in the machining process generation has been carried out at two stages starting from the representation of the CAD model of die by STL facets. In the first stage the difficulty linked to the variation of cutting speed contact areas between cutting tools and the shape of the die have been analysed. Knowledge formalization at this stage leads to the generation of contact area map which is based on the orientation of the normal vector of each facet of the STL model compared to the tool axis direction. The decomposition of the normal vector by projection respectively to the tool axis direction and the plane H perpendicular to this tool axis direction defines the orientation indicator and the contact area indicator. These two indicators are useful for the integration of knowledge associated to cutting tools. The second stage of knowledge formalization is based on the difficulty to keep the orientation of the normal vectors of several facets in the same plane. When a plane parallel to the tool axis direction exists, it defines a machining continuity and corresponds to the basic plane for the parallel planes strategy. Maps generated at this stage for the search of continuity planes are based on a continuity indicator defined according to a given machining direction. The continuity indicator provided information for the integration of knowledge associated to machining strategies. Extraction of geometrical features from the combination of the two types of maps is not automatic. This is one of the points which must be developed in next work. The knowledge integration being based on several expert rules, the geometric feature extraction process must be based on artificial intelligence approaches like neural network.

The industrial case study presented in the final part of the paper shown a complete application of the knowledge formalization for the decomposition of the die shape into geometrical features. The machining features created by associating cuttings tools and machining strategies to the geometrical features can be used by a machining assistant to generate the machining sequences.

## Web references